\documentstyle[12pt,epsfig]{article}

\begin{document}

\begin{center}

{\Large Mass scale effects for the Sudakov form factors in theories 
with the broken gauge symmetry}

\vspace*{0.3 cm}

{\large A.~Barroso and   
B.I.~Ermolaev\footnote{Permanent address:
 A.F.~Ioffe Physico-Technical Institute, 
 St.Petersburg 194021, Russia}\\
Centro de F\'{\i}sica Te\'orica e Computacional, Faculdade de Ci\^encias, Universidade de Lisboa, Av. Prof. Gama Pinto 2,
P-1649-003 Lisboa, Portugal\\}

\end{center}

\begin{abstract}
The off-shell and the on-shell 
Sudakov form factors in theories with broken gauge symmetry 
are calculated in the double-logarithmic approximation.
We have used different infrared cut-offs, i.e. different mass scales, for
virtual photons and weak gauge bosons.
  
\end{abstract}

\section{Introduction}

In QED the electro-magnetic vertex function, $\Gamma_{\mu}$, can be
written as:   
\begin{equation}
\label{gammaborn}
\Gamma_{\mu} = \bar{u}(p_2)\Big[ \gamma_{\mu}  f(p_1,~p_2) -  
(1/2m)\sigma_{\mu \nu}  q_{\nu} g(p_1, p_2) \Big] u(p_1) ~, 
\end{equation}
where $f$ and $g$ are the form factors.
In the fifties V.V.~Sudakov showed\cite{sud} that 
in the limit of large momentum transfer, i.e.,
\begin{equation}
\label{kin}
q^2 = (p_2 - p_1)^2 \gg p^2_1, p^2_2 ~,
\end{equation}
the most important radiative corrections to the form factor 
$f(p_1, ~p_2)$ are the double- 
logarithmic ones (DL). The summing of these corrections to all orders in $\alpha$ 
leads to,  
\begin{equation}
\label{foffshell}
f = e^{-(\alpha/2 \pi) \ln (q^2/p^2_1)\ln(q^2/p^2_2)},
\end{equation} 
for off-shell momenta $p_1$ and ~ $p_2$ and to the formula 
\begin{equation}
\label{fonshell}
  f = e^{-(\alpha/4 \pi) \ln^2 (q^2/m^2)},
\end{equation} 
if $p^2_1 = p^2_2 = m^2$.

After this pioneer work, the Sudakov form factor $f$ was calculated in QCD  
\cite{qcd} and recently it has also been considered in 
the electroweak (EW) theory \cite{flmm},\cite{ciaf},\cite{k}. 
In non-Abelian theories the 
 direct graph-by-graph calculation to all orders in the couplings is  
a very complicated procedure, 
even when the double- logarithmic approximation (DLA) is used.  
Technically, 
it is more convenient to use some evolution equation.  
 In particular, the infrared evolution equation 
(IREE) approach was used in ref. \cite{flmm} to calculate the ``inclusive''   
EW Sudakov form factor, i.e., where summation over the left handed lepton
flavour was assumed.
The IREE method is based on the Gribov bremsstrahlung theorem \cite{g}. 
It was applied earlier, 
in ref. \cite{efl}, to calculate the radiative form factors for 
$e^+e^-$ annihilation into quarks and gluons. It has been proved to
be a very efficient and simple method. 
This theorem was formulated and proved first for 
QED and then generalised in refs.\cite{kl},\cite{efl},\cite{e},\cite{ce} to 
QCD. 
Besides the calculation of the Sudakov form factor $f(q^2)$, the IREE
turned out to be also useful \cite{et} in order to calculate 
 the form factor $g(q^2)$ of electrons and quarks 
in the kinematic region specified by eq.(\ref{kin}).

The IREE exploits the evolution of 
the scattering amplitudes with respect to the 
infrared cut-off $\mu$ introduced in the space of the  
transverse momenta of virtual particles.  
This cut-off plays the role of 
a mass scale and with DL accuracy all other masses can be safely
neglected.   
So, in theories with unbroken gauge symmetry 
it is enough to have one mass scale. 
On the other hand, for the electroweak theory, 
the $SU(2)\times U(1)$ symmetry is broken down to the $U_{EM}(1)$
symmetry. 
This introduces a second mass parameter, $M$, associated with the
symmetry breaking scale.  Therefore, besides the conventional 
DL contributions of the order of $\ln^{2n}(q^2/M^2)$ in $n$-th order of the 
perturbative expansion, there appear other corrections  of the type
\begin{equation}
\label{c}
\sim a_k \ln^k (M^2/ \mu^2) \ln^{(n - k)}(q^2/M^2) ,
\end{equation}
where $a_k$ are numerical coefficients 
and $k$ runs from 1 to $n$. 
Since $\mu$ and $M$ can be widely different 
(we assume the value of $\mu$  to be equal or 
greater than masses of 
the involved fermions whereas $M$ is comparable with masses of the 
weak bosons), the impact of 
 these two mass scales on the value of the inclusive  Sudakov form 
factor  can be important. To calculate these contributions to all orders 
in the electroweak  couplings is the aim of the present work.  
In Sect. 2 we obtain the Sudakov form factor for a $U(1)\times U(1)$ model 
with two 
``photons'', using the graph-by-graph calculation. 
In Sect. 3 we derive and solve the IREE for the Sudakov form factor
for the same model. 
Then, in Sect. 4 a similar derivation is made 
for the Sudakov form factor in the 
electroweak theory. 
Expressions for the off-shell electroweak Sudakov form factor are 
obtained in Sect.~5.  
Finally, in Sect.~6 we summarise and discuss our results.

\section{The Sudakov form factor in the Abelian model}

As a toy model it is instructive to consider a $U(1)\times U(1)$ gauge 
theory. The first group is the normal electro-magnetic gauge group and
the second $U(1)$ is broken. The corresponding gauge boson, called $B$
meson has a mass $M$. Besides the photon and the $B$ meson the model
has only one charged fermion, and a neutral scalar Higgs particle
arising from the spontaneous breaking of the second $U(1)$. 

\begin{figure}[htbp]
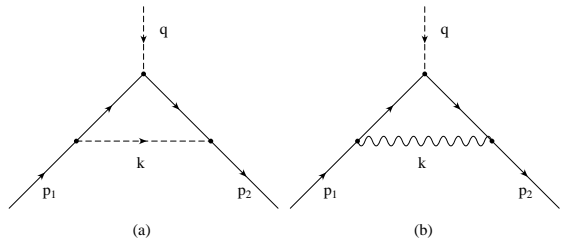

  \begin{center}
    \epsfig{file=figures/Diag1a.epsi,width=3.6cm}
    \epsfig{file=figures/Diag1b.epsi,width=3.6cm}
   \caption{Feynman diagrams that give the Sudakov form factors in
     first order}
    \label{fig1}
  \end{center}
\end{figure}

At one loop order the Sudakov form factor is obtained from the
diagrams in figure \ref{fig1} where the dashed line represents the
photon and the wavy line represents the $B$ meson.
Summing both contributions in $DL$ approximation the result is
\begin{equation}
\label{f1}
f^{(1)} = -(g_1^2/16 \pi^2) \ln^2(q^2/ \mu^2) - 
(g_2^2/16 \pi^2) \ln^2(q^2/ M^2)
\end{equation}
where $g_1, g_2$ are the gauge couplings corresponding to the unbroken 
and broken groups respectively. It is interesting to point out that
there is also a similar diagram with the higgs particle in loop.
However this contribution vanishes as $(m/M)^2$ when the fermion mass, 
$m$, goes to zero.
This is obviously true in all orders of perturbation theory.
 
\begin{figure}[htbp]
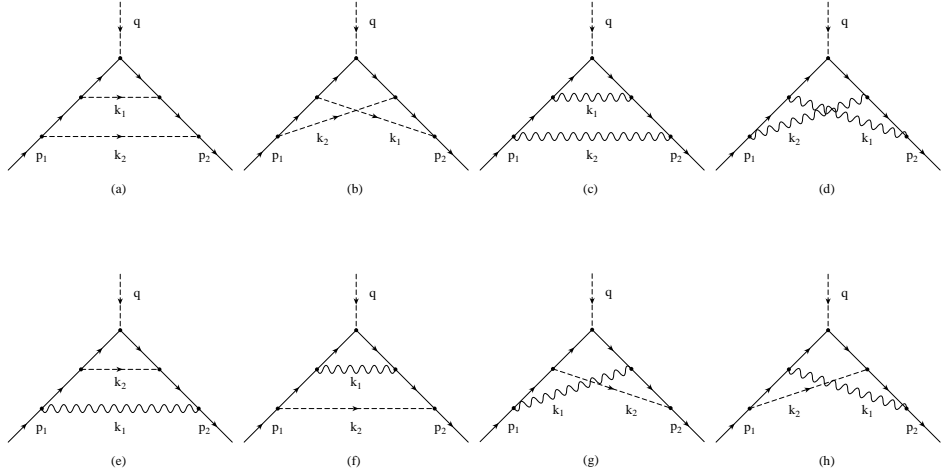

  \begin{center}
    \epsfig{file=figures/Diag2a.epsi,width=3.cm}
    \epsfig{file=figures/Diag2b.epsi,width=3.cm}
    \epsfig{file=figures/Diag2c.epsi,width=3.cm}
    \epsfig{file=figures/Diag2d.epsi,width=3.cm}\\[1cm] 
    \epsfig{file=figures/Diag2e.epsi,width=3.cm}
    \epsfig{file=figures/Diag2f.epsi,width=3.cm}
    \epsfig{file=figures/Diag2g.epsi,width=3.cm}
    \epsfig{file=figures/Diag2h.epsi,width=3.cm}
   \caption{Feynman diagrams that give the Sudakov form factors in
     second order}
    \label{fig2}
  \end{center}
\end{figure}

\begin{figure}[htbp]
  \begin{center}
    \epsfig{file=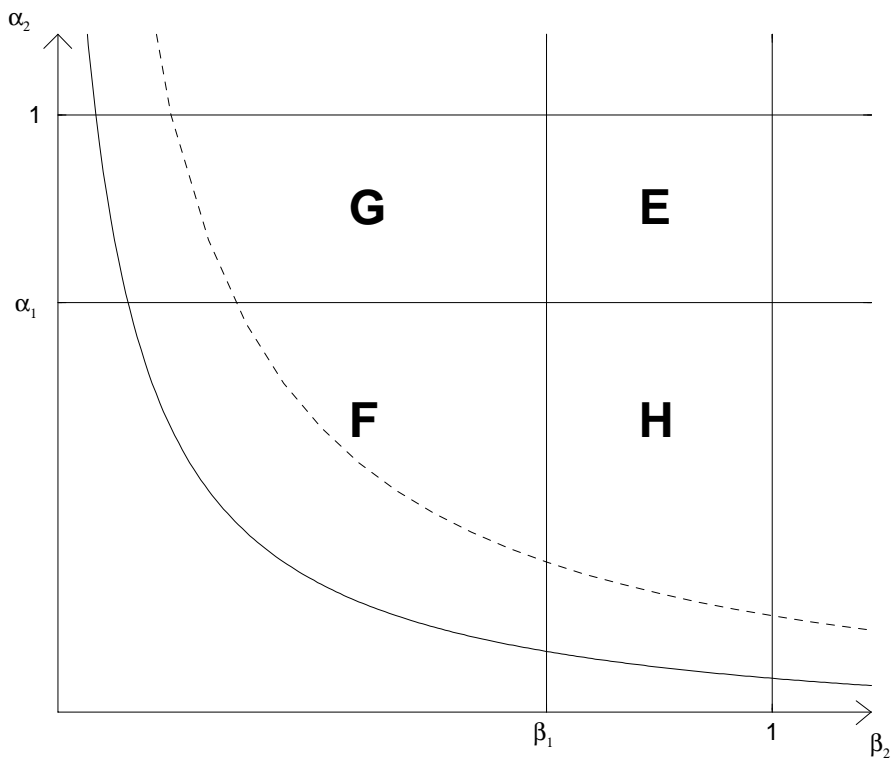,width=10cm,angle=-90}
   \caption{Integration regions in the $\alpha_2$ $\beta_2$ plane.}
    \label{fig3}
  \end{center}
\end{figure}

At two loop order the DL contributions stems  from the diagrams of
figure Fig.~\ref{fig2}. 
The calculation of the first four diagrams in Fig.\ref{fig2} is
similar to the QED calculation.
Hence, we simply quote the result,
\begin{equation}
\label{ab}
f^{(2)}_{a+b+c+d} = \frac{1}{2}[(g_1^2/16 \pi^2) \ln^2(q^2/ \mu^2)]^2
+\frac{1}{2}[(g_2^2/16 \pi^2)\ln^2(q^2/ M^2)]^2.
\end{equation}
The calculation of the remaining four diagrams is less trivial. Let
us consider diagram $e)$, for instance. It is well known that in DLA
one can perform the integration over the momenta $k_i$, $i=1,2$,
replacing the boson propagators by $-2\pi i \delta(k_i^2)$. Then,
using the Sudakov parametrisation, 
\begin{equation}
\label{sud}
k_i = \alpha_i p_2 + \beta_i p_1 + k_{i\perp},
\end{equation}
it is easy to integrate over $k_{i\perp}$ and to obtain 
\begin{equation}
\label{e}
f_e^{(2)} = \frac{g^2_1 g^2_2}{(8 \pi^2)^2} \int_{D_e}
\frac{d \alpha_1 d \beta_1 d \alpha_2 d \beta_2}
{\alpha_1 \beta_1 \alpha_2 \beta_2} 
\Theta (\alpha_1\beta_1 - \lambda^2_1)~ 
\Theta(\alpha_2\beta_2 - \lambda^2_2), 
\end{equation} 
where $\lambda^2_1=M^2/s$, $\lambda^2_2=\mu^2/s$, and  
$s=|q^2|$.  
Notice that the $DL$ arises if one uses the approximation $\alpha_1 +
\alpha_2 \approx \alpha_2$ and $\beta_1+ \beta_2 \approx \beta_2$.
This implies $\alpha_2 \gg \alpha_1$ and $\beta_2 \gg \beta_1$. These
conditions plus the boundaries stemming from the arguments of the $\Theta$ 
functions define the integration region, $D_e$.
In Fig. \ref{fig3} we show this region.
The full curve corresponds to the equation
$\alpha_2\beta_2=\lambda_2^2$ and the dashed curve represents the
condition $\alpha_1\beta_1=\lambda_1^2$. Because the lower limits of
the $\alpha_2$ and $\beta_2$ integrals are $\alpha_1$ and $\beta_1$,
respectively, it is easy to see that we are integrating over the
rectangle $E$.
So we obtain
\begin{equation}
\label{e}
f_e^{(2)} = \frac{g^2_1}{8 \pi^2}\frac{g^2_2}{8 \pi^2}
\int_{\lambda_1^2}^1\frac{d\alpha_1}{\alpha_1} 
\int_{\lambda_1^2/\alpha_1}^1\frac{d\beta_1}{\beta_1}
\int_{\alpha_1}^1\frac{d\alpha_2}{\alpha_2} 
\int_{\beta_1}^1\frac{d\beta_2}{\beta_2}.
\end{equation} 
The integrals are now trivial. However, rather than doing this it is
better to realize that the remaining diagrams give rise to similar
integrals but with the integration regions $F$, $G$ and $H$ of
Fig. \ref{fig3}, respectively. Then, the sum of the diagrams $e)$,
$f)$, $g)$ and $h$ corresponds to 
\begin{equation}
f_{e+f+g+h}^{(2)} = \frac{g^2_1}{8 \pi^2}\frac{g^2_2}{8 \pi^2}
\int_{\lambda_1^2}^1\frac{d\alpha_1}{\alpha_1} 
\int_{\lambda_1^2/\alpha_1}^1\frac{d\beta_1}{\beta_1}
\int_{\lambda_2^2}^1\frac{d\alpha_2}{\alpha_2} 
\int_{\lambda_2^2/\alpha_2}^1\frac{d\beta_2}{\beta_2}.
\end{equation} 
which immediately leads to 
\begin{equation}
\label{eh}
f_{e+f+g+h}^{(2)} = \frac{g_1^2}{16 \pi^2} \ln^2(q^2/ \mu^2)
\frac{g_2^2}{16 \pi^2} \ln^2(q^2/ M^2)~.
\end{equation} 

Adding these results to eq.(\ref{ab}) one obtains the total two-loop
contribution to the form factor, namely:
\begin{equation}
\label{ah}
f^{(2)} = (1/2)\big[(g_1^2/16 \pi^2) \ln^2(q^2/ \mu^2) + 
(g_2^2/16 \pi^2) \ln^2(q^2/ M^2) \big]^2~.
\end{equation} 
Repeating the same analyses in higher orders in $g_1$ and $g_2$, we
would arrive at the simple exponentiation of the double-logarithmic
corrections to $\Gamma_{\mu}$, i.e., 
\begin{equation}
\label{fmum}
\Gamma_{\mu}^{DL} = \Gamma_{\mu}^{Born} f(q^2, \mu^2, m^2)~, 
\end{equation}
with 
\begin{equation}
\label{exp}
f(q^2, \mu^2, M^2) = 
e^{- \big[(g_1^2/16 \pi^2) \ln^2(q^2/  \mu^2) + 
(g_2^2/16 \pi^2) \ln^2(q^2/ M^2) \big]} .
\end{equation}
This equation accounts for all DL contributions  described by  
Eq.~(\ref{c}).

\section{Infrared evolution equations with two mass scales}

In order to avoid the direct graph-by-graph summation of the DL contributions, 
it is possible to obtain the Sudakov form factor 
$f(q^2, \mu^2, m^2)$ as a solution of some integral equation. 
The method of obtaining this infrared evolution 
equation (IREE) can be extended in order to include two mass scales,
$\mu$ and $M$.

\begin{figure}[htbp]
  \begin{center}
    \epsfig{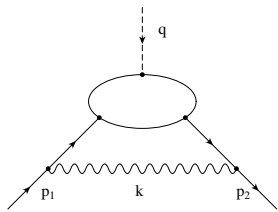}
   \caption{Factorisation of boson with the minimal $k_{\perp}$.}
    \label{fig4}
  \end{center}
\end{figure}

Let us first notice that the boson mass $M$ in virtual propagators 
can be regarded in DLA as the infrared cut-off for integrating over
transverse momentum space  for  $B$-bosons. In fact, with logarithmic
accuracy we have
\begin{equation}
\label{ircutoff}
\int_0^s \frac{d k^2_{\perp}}{k^2_{\perp} + M^2} =  
\int_{M^2}^s\frac{d k^2_{\perp}}{k^2_{\perp}} ~.
\end{equation}
In DLA the integrals over the longitudinal momenta have the transverse 
momenta as the lowest limit. So, after introducing the cut-off shown in 
the previous equation one can neglect the mass $M$ in the
$B$-propagators and still be free of infra-red singularities.   
On the other hand,  
$\mu$ is the IR cut-off in the transverse momentum space for 
photons. Therefore, with the DL 
accuracy, we have QED with two kinds of photons, each one has an independent  
infrared cut-off. We remind the reader that we have 
assumed that $\mu \ll M$. 
According to 
the generalisation\cite{efl,ce} of the Gribov bremsstrahlung 
theorem\cite{g}, the boson with the minimal $k_{\perp}$ can be factorized, 
i.e. the main contribution from integrating over its momentum comes from the 
graphs where its propagator is attached, in all possible ways, 
to the external charged lines whereas 
its $k_{\perp}$ acts as a new IR cut-off for the integrations over the
remaining loop momenta and becomes, in DLA, a new effective mass scale.
Therefore the  blob in Fig.\ref{fig4} is on-shell. 
It depends on the new 
infrared cut-off $k_{\perp}$  and does not 
depend on the longitudinal components of momentum $k$. However, in contrast 
to the usual QED situation, now we have two options: 
the factorized particle with minimal $k_{\perp}$ can be 
either a photon or a $B$-boson.     
Applying the Feynman rules to the graph in Fig.\ref{fig4}, we obtain
for the first possibility 
\begin{eqnarray}
\label{photon}
M^{\gamma} = - \frac{g_1^2}{8 \pi^2} 
\Big[ \int_{\mu^2}^{M^2} \frac{d k^2_{\perp}}{k^2_{\perp}} 
\ln(s / k^2_{\perp}) f(s/ k^2_{\perp},~s/M^2) +  \\  \nonumber 
\int_{M^2}^{s} \frac{d k^2_{\perp}}{k^2_{\perp}} 
\ln(s / k^2_{\perp}) f(s/ k^2_{\perp},~s/k^2_{\perp}) \Big]~, 
\end{eqnarray}   
where $\ln(s/ k^2_{\perp})$ appears as a result of integrating over the 
longitudinal momentum of the factorized photon quite similarly to the way 
it appears in the first loop calculation of $f$ in QED.
Note that the form factors $f$ in the right hand side (rhs) of
eq.~(\ref{photon}) have the same first argument, $s/k^2_{\perp}$, but 
different second arguments, $s/M^2$,   in the first integral 
and  $s/k^2_{\perp}$ in the second one. The reason for this is that the   
propagators of the $B$-bosons depend on  $M$ through  $(k^2_{\perp} +
M^2)$ and in DLA they are approximated by $max(k^2_{\perp}, ~M^2)$ . 
The second possibility is that the factorized particle is the $B$-boson. 
Now, we can regard $M$ as the infra-red cut-off for the integration
over the $k_{\perp}$ of the $B$-meson.
Hence the result is 
\begin{equation}
\label{boson}
M^{B} = - \frac{g_2^2}{8 \pi^2}
\int_{M^2}^{s} \frac{d k^2_{\perp}}{k^2_{\perp}} 
\ln(s / k^2_{\perp}) f(s/ k^2_{\perp},~s/k^2_{\perp}) .
\end{equation}

Adding eqs.~(\ref{photon}) and (\ref{boson}) and including the Born 
contribution, $f_{Born} = 1$, one 
obtains the integral IREE for 
the form factor $f(s/ \mu^2,~ s/m^2)$, i.e.
\begin{equation}
\label{eqf}
f(s/ \mu^2,~ s/M^2) = 1 + M^{\gamma} + M^{B} ~.
\end{equation}
The solution of Eq.~(\ref{eqf}) can be obtained, for example, 
by iterations: substituting $f_{Born} = 1$ into the integrand of the rhs 
of  eq.~(\ref{eqf}) we obtain $f(s/ \mu^2,~ s/m^2)$ in 
the one loop approximation and so on. It is easy to see that, after 
summing up contributions to all orders,   
the solution to eq.~(\ref{eqf}) coincides with 
Eq.~(\ref{exp}) obtained by the direct graph-by-graph summation. \\ 
Alternatively, the integral IREE given in eq. (\ref{eqf}) can be 
rewritten in the differential form (cf \cite{flmm}). Indeed, 
differentiating Eq.~(\ref{eqf}) with respect to  
$\mu$ yields
\begin{equation}
\label{dmu}
\frac{\partial f(\rho_1,~\rho_2)}{\partial \rho_1} = 
- \frac{g_1^2}{8 \pi^2} \rho_1 f(\rho_1,~\rho_2) 
\end{equation}
where $\rho_1 = \ln(s/\mu^2)$ and $\rho_2 = \ln(s/M^2)$. Obviously the
solution is
\begin{equation}
\label{fmu}
f(s/ \mu^2,~ s/M^2) = G(\rho_2) e^{ -g_1^2/ 16 \pi^2 \rho_1^2}.
\end{equation}
Substituting it into Eq.~(\ref{dmu}) and differentiating  with 
respect to  $\rho_2$ gives 
\begin{equation}
\label{dm}
\frac{d G(\rho_2)}{d \rho_2} = 
- \frac{g_2^2}{8 \pi^2} \rho_2 G(\rho_2)~. 
\end{equation}
With the boundary condition $G(0) = 1$ it is easy to solve this
 equation. 
Again one obtains eq.(\ref{exp})

\section{The Sudakov form factor in the electroweak theory} 
The fermions in the electroweak theory are such that the left-handed
fields are doublets of the weak isospin group and the right handed fields
are singlets of $SU(2)\times U_y(1)$. So one has a left, $F_L$, and a
right, $F_R$, Sudakov form factors. In line with the approximation of
massless particles there is no chirality flip amplitude. Because the
right-handed fermions couple to the $U_Y(1)$ boson, $F_R$ only
gets contributions from $Z$ and photon exchange. Hence, borrowing
directly from the results of the previous section, one can easily obtain:
\begin{equation}
\label{fr}
F_R(s/ \mu^2, ~s/M^2) =\exp(- \psi_R)~,
\end{equation} 
with 
\begin{eqnarray}
\label{psir}
\psi_R = 
\frac{\alpha Q^2}{4 \pi}
[\ln^2(s/\mu^2) + \tan^2 \theta \ln^2(s/M^2)]~, 
\end{eqnarray}  
where $M=M_Z$ and $\theta_W$ is the Weinberg angle.

Now, lets us derive the IREE for $F_L(s/\mu^2,s/M^2)$, neglecting the
mass difference between the $Z$ and the $W$ boson, i.e., $M_W=M_Z=M$.
To do this, applying the Gribov bremsstrahlung theorem, one should
factorize the boson with the minimal $k_\perp$. If the integration
over the $k_\perp$ of the factorized boson is done in the region
$M<k_\perp<\sqrt{s}$ this particle can be any of the four bosons
present in the theory, the $W^\pm$, the $Z$ and the photon. This
yields the following contribution to the IREE:
\begin{equation}
\label{wz}
\tilde{M}^{WZ} = - \frac{e^2Q^2 + C_{WZ}}{8 \pi^2}
\int_{M^2}^{s} \frac{d k^2_{\perp}}{k^2_{\perp}} 
\ln(s / k^2_{\perp}) F(s/ k^2_{\perp},~s/k^2_{\perp}) ,
\end{equation}
where we have explicitly separated the photon from the $WZ$
contributions. The latter are proportional to 
\begin{equation}
\label{cwz}
C_{WZ} = g^2 [(t^2_1 + t^2_2) + 
(1/ \cos^2 \theta_W)(t_3 - \sin^2 \theta_W Q)^2].
\end{equation}
The second DL region is when $\mu<k_\perp<M$, but, now, the
integration over the $k_\perp$ of the factorized boson only gives a DL
contribution if this boson is a photon. So we obtain 
\begin{equation}
\label{phot}
\tilde{M}^{\gamma} = - \frac{e^2 Q^2}{8 \pi^2} 
\int_{\mu^2}^{M^2} \frac{d k^2_{\perp}}{k^2_{\perp}} 
\ln(s / k^2_{\perp}) F(s/ k^2_{\perp},~s/M^2)~. 
\end{equation}   

The sum of the factorized contributions given by eqs.(\ref{wz}) and
(\ref{phot}) together with the Born value $F_{Born} = 1$ leads to
the IREE for the Sudakov form factor $F_L(s/ \mu^2,~s/M^2)$ 
in the integral form,

\begin{eqnarray}
\label{eqF}
F(s/ \mu^2,~s/M^2) = 1 
- \frac{e^2Q^2}{8 \pi^2} 
\int_{\mu^2}^{M^2} \frac{d k^2_{\perp}}{k^2_{\perp}} 
\ln(s / k^2_{\perp}) F(s/ k^2_{\perp},~s/M^2) -  \nonumber \\ 
- \frac{(C_{WZ} + e^2 Q^2)}{8 \pi^2}
\int_{M^2}^{s} \frac{d k^2_{\perp}}{k^2_{\perp}} 
\ln(s / k^2_{\perp}) F(s/ k^2_{\perp},~s/k^2_{\perp}).  
\end{eqnarray}   
This equation is similar to Eq.~(\ref{eqf}) 
where $g_1$ is replaced by $e Q$ and $g_2$ is 
replaced by 
$C_{WZ}$~. 
Differentiating Eq.~(\ref{eqF}) first with respect to $\mu$ and then 
with respect to $M$ we obtain the IREE in the 
differential form (cf  eqs.~(\ref{dmu}),(\ref{dm})):

\begin{equation}
\label{dmuF}
\frac{\partial F_L(\rho_1,~\rho_2)}{\partial \rho_1} = 
- \frac{e^2 Q^2}{8 \pi^2} \rho_1 F_L(\rho_1,~\rho_2) ~,
\end{equation} 

\begin{equation}
\label{dmF}
\frac{d \tilde{G}(\rho_2)}{d \rho_2} = 
- \frac{C_{WZ}}{8 \pi^2} \rho_2 \tilde{G}(\rho_2)~,
\end{equation}
where $\tilde{G}$ is the general solution of
eq.~(\ref{dmuF}). Finally, solving this equation in the same way that
we have solved  eq.~(\ref{dm}) we obtain 
\begin{equation}
\label{F}
F_L(s/ \mu^2, ~s/M^2) = \exp(-\Psi_L) ~,
\end{equation}      
with
\begin{eqnarray}
\label{psi}
\Psi_L &=& \frac{e^2 Q^2}{16 \pi^2} \ln^2(s/ \mu^2)  \nonumber \\
& & + \frac{g^2}{16 \pi^2}
[(t^2_1 + t^2_2) + \frac{1}{\cos^2 \theta_W}(t_3 - \sin^2 \theta_W Q)^2]
\ln^2(s/ M^2) ~.
\end{eqnarray}

\section{The off-shell Sudakov electroweak form factor }

Eqs.~(\ref{foffshell}) and (\ref{fonshell}) show that even in the
simplest QED  case the on-shell Sudakov form factor cannot be obtained
from the  expression for the off-shell form factor with the simple
replacement  of the electron $p^2_1$ and  $p^2_2$ by the electron mass
or by the infrared cut-off. 
Both form factors have to be calculated independently. 
Obviously, the same is true for the off-shell electroweak  
Sudakov form factor $\tilde{F}_L$. 
Before we calculate $\tilde{F}_L$, it is instructive to demonstrate
how the IREE for the off-shell QED form factor can be obtained. 
This form factor $\tilde{f}$ depends on $q^2,~p^2_1,~p^2_2$ and
also depend  on the infrared cut-off $\mu$, i.e., $\tilde{f} =
\tilde{f}((q^2/ \mu^2,~p^2_1/\mu^2, ~p^2_2/\mu^2 )$.    

\begin{figure}[htbp]
  \begin{center}
    \epsfig{file=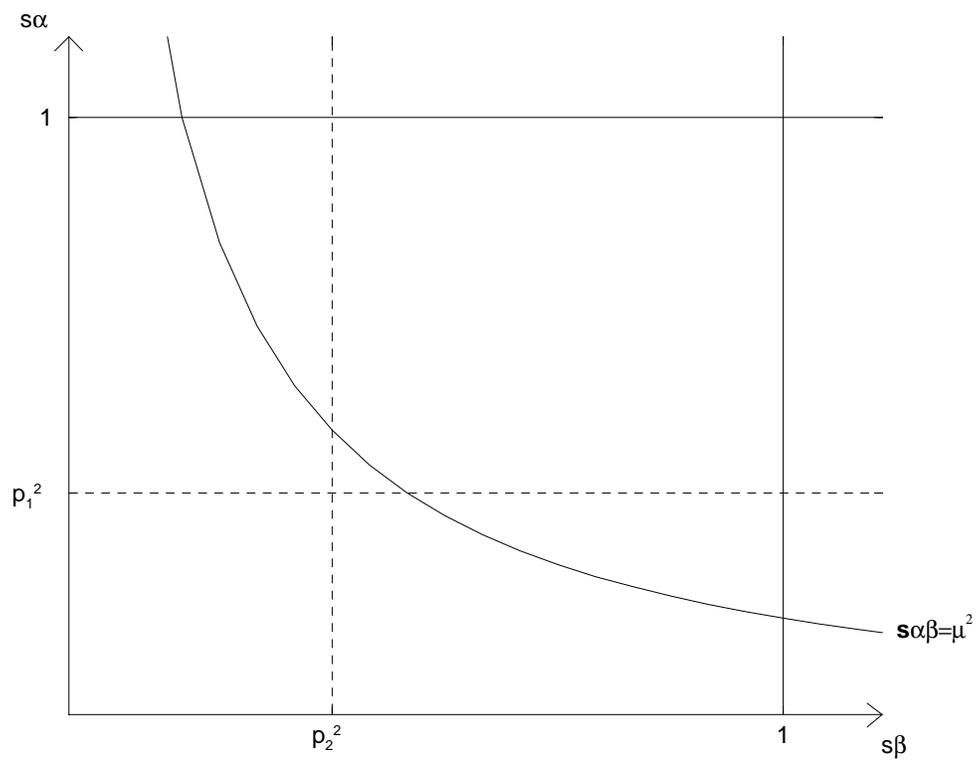,width=10cm,angle=-90}
   \caption{The integration region $D_1$}
    \label{fig5}
  \end{center}
\end{figure}

Similarly to the on-shell case, factorizing the contribution of the
virtual photon with the minimal  $k_{\perp}$ leads to the following
IREE:  
\begin{equation}
\label{offint}
\tilde{f}((q^2/ \mu^2,~p^2_1/\mu^2, ~p^2_2/\mu^2 ) = 1 -  
\frac{\alpha}{2 \pi}
\int_D \frac{d k^2_{\perp}}{k^2_{\perp}}\frac{d \beta}{\beta}
\tilde{f}((q^2/ k^2_{\perp},~p^2_1/k^2_{\perp}, ~p^2_2/k^2_{\perp} )~.
\end{equation}
However, in contrast to the IREE for the on-shell form factor, the 
region $D$ in the previous equation now depends on $p^2_1$ and $p^2_2$.  
The region  $D$ is different for the particular case when 
the virtualities $p^2_{1,2}$ are small enough such that  
\begin{equation}
\label{d1}
p^2_1 p^2_2 < q^2 \mu^2
\end{equation}  
or large enough so that  
\begin{equation}
\label{d2}
p^2_1p^2_2 > q^2 \mu^2 ~.
\end{equation} 
In the former case we denote the integrating region $D_1$ and call it
$D_2$ in the latter case.
In Fig. \ref{fig5} we show $D_1$.
 In the plane $k_\perp^2=\alpha\beta$, $\beta$, 
 $D_1$ is bounded by the ``on-shell'' curves
$s\beta=k_\perp^2$ and $s\beta=s$ and also by the new curves
$s\beta=p_2^2$ and $p_1^2p_2^2=q^2\mu^2$. From now on we call
$\tilde{f}_1$ the contribution to $\tilde{f}$ from the region $D_1$.
Performing the $\beta$ integration in eq.(\ref{offint}) and after
that differentiating with respect to $\mu^2$ we obtain 
\begin{equation}
\label{offdif} 
\frac{\partial \tilde{f}_1}{\partial x} + 
\frac{\partial \tilde{f}_1}{z_1} + \frac{\partial \tilde{f}_1}{ z_2} = 
- \frac{\alpha}{2 \pi}\big[ x - z_1 - z_2 \big] \tilde{f}_1  ,
\end{equation}
where $x = \ln(q^2/ \mu^2)$, $~z_1 = \ln(p^2_1/\mu^2)$ and 
$~z_2 = \ln(~p^2_2/\mu^2)$.
Obviously, the solution to this equation is,
\begin{equation}
\label{offsol1}
 \tilde{f}_1(q^2,p^2_1,p^2_2, \mu^2) = 
e^{-\big[(\alpha/4 \pi)(x^2 - z^2_1 - z^2_2) \big]}~.
\end{equation}

In the kinematical region specified by eq.(\ref{d2}), the integration
of eq.(\ref{offint}) is over the domain $D_2$. This region does not
involve $\mu$ so we have 
\begin{equation}
\label{eqd2}
-\mu^2\frac{\partial \tilde{f}_2}{\partial \mu^2} = 
\frac{\partial \tilde{f}_2}{\partial x} + 
\frac{\partial \tilde{f}_2}{z_1} + \frac{\partial \tilde{f}_2}{ z_2} = 0
\end{equation}
and its general solution is 
\begin{equation}
\label{d2sol}
\tilde{f}_2 = \Phi(x - z_2, x - z_2)
\end{equation} 
where $\Phi$ is an arbitrary function. The matching condition
$\tilde{f}_1 = \tilde{f}_2$ when $p^2_1p^2_2 = q^2 \mu^2 $, which is
equivalent to $x=z_1+z_2$, leads to 
\begin{equation}
\tilde{f}_2 = \exp \big[
-(\alpha/2\pi)\ln(q^2/p^2_1)\ln(q^2/p^2_2)\big].
\end{equation}
This is exactly the expression that we had anticipated in the
introduction (cf. eq.(\ref{foffshell})).

Now, it should be clear that following a similar prescription one
obtains the IREE for $\tilde{F}_L$, namely
\begin{eqnarray}
\label{eqFo}
\tilde{F}(s/ \mu^2,~s/M^2) = 1 
- \frac{e^2}{8 \pi^2} 
\int_{D} \frac{d k^2_{\perp}}{k^2_{\perp}} 
\ln(s / k^2_{\perp}) \tilde{F}(s/ k^2_{\perp},~s/M^2) -  \nonumber\\ 
- \frac{(C_{WZ} + e^2 Q^2)}{8 \pi^2}
\int_{D'} \frac{d k^2_{\perp}}{k^2_{\perp}} 
\ln(s / k^2_{\perp}) \tilde{F}(s/ k^2_{\perp},~s/k^2_{\perp}).  
\end{eqnarray}   
The regions $D$ and $D'$ are bounded, in addition to the ``on-shell'' 
requirements, by the relation between $q^2 \mu^2$, $q^2 M^2 $ and
$p^2_1p^2_2$.  
We specify the following basic off-shell kinematic regions: \\ \\
$R_1$:~$\mu^2<p^2_1,p^2_2 < M^2,~~~~~p^2_1p^2_2 <q^2 \mu^2 $,\\ \\
$R_2$:~$\mu^2<p^2_1,p^2_2 < M^2,~~~~~q^2 \mu^2 < p^2_1p^2_2  < q^2 M^2
$ \\ \\
$R_3$: $p^2_1p^2_2 > M^2,~~~~~~p^2_1p^2_2 <q^2 M^2 $.\\ \\
$R_3$: $p^2_1p^2_2 > M^2,~~~~~~p^2_1p^2_2 >q^2 M^2 $. \\ \\ \\
For each region, $R_i$ ($i=1\dots 4$), we obtain
\begin{equation}
\label{offF}
\tilde{F_L} = \exp(-\psi_i),
\end{equation} 
with
\begin{equation}
\label{psi1}
\psi_1 = \frac{e^2 Q^2}{16 \pi^2} 
[\ln^2(q^2/ \mu^2) - \ln^2(p^2_1/\mu^2) - \ln^2(p^2_2/\mu^2)] 
 + \frac{g^2}{16 \pi^2}
C_{WZ}\ln^2(s/ M^2) ~
\end{equation}
for the region $R_1$, 
\begin{equation}
\label{psi2}
\psi_2 = \frac{e^2 Q^2}{8 \pi^2} \ln(q^2/ p^2_1)\ln(q^2/ p^2_2)   
 + \frac{g^2}{16 \pi^2}
C_{WZ}\ln^2(s/ M^2) ~
\end{equation}
for the region $R_2$, 
\begin{equation}
\label{psi3}
\psi_3 = \frac{e^2 Q^2}{8 \pi^2} \ln(q^2/ p^2_1)\ln(q^2/ p^2_2)   
 + \frac{g^2}{16 \pi^2}
C_{WZ}[\ln^2(q^2/ M^2) - \ln^2(p^2_1/M^2) - \ln^2(p^2_2/M^2)] ~
\end{equation} 
for the region $R_3$, and finally 
\begin{equation}
\label{psi3}
\psi_4 = \frac{e^2 Q^2 + C_{WZ}}{8 \pi^2} \ln(q^2/ p^2_1)\ln(q^2/ p^2_2) 
\end{equation}  
for the region $R_4$.

\section{Discussion}

Expression ~(\ref{F}) for the electroweak Sudakov form factor, $F_L$,
accounts for the mass difference between the photon and the weak
bosons.
On the other hand, it neglects the difference between the masses of
the $W$ and the $Z$.
It has been obtained introducing different infrared cut-offs for the
integration over transverse momenta of different gauge bosons: cut-off
$\mu$ for photons and cut-off $M$ for the $W$ and the $Z$. 
Expanding Eq.~(\ref{F}) into serias, one can easily extract the 
first-loop and the second-loop DL contributions. The first-loop 
contribution (save the minus sign) is given by Eq.~(\ref{psi}). 
The DL contributions to $F_L$ in two loops   
were also calculated in \cite{bw}. Before comparing our results with 
results of \cite{bw}, let us notice that besides the DL contributions 
we account for, the DL contributions in 
\cite{bw} account also for the double logarithms 
of $m^2/\mu^2$, where $m$ stands 
for fermion masses. Such contributions are absent if the photon cut-off 
is equal or greater than masses of involved fermions as we assume in 
this paper. Having dropped them, we arrive at agreement with the 
two-loop results of \cite{bw}.  
Usually, DL calculations involve only one mass scale. Using one mass
scale, for instance $M$, for all DL terms in $\Psi_L$ of
Eq.~(\ref{psi}) allows us to rewrite it as  
\begin{eqnarray}
\label{onescale}
 \Psi_L = \frac{g^2}{16 \pi^2}
(t^2_i + g'^2(Y/2)^2)\ln^2(s/M^2) + 
\frac{e^2 Q^2}{16 \pi^2} [2\ln(s/ M^2) \ln(M^2/\mu^2) \nonumber \\ 
 + \ln^2(M^2/\mu^2)].~~~
\end{eqnarray}
The first term in the rhs of this equation is the DL contribution with
the same scale $M$ for both the electro-magnetic and the weak
interactions. The second term is formally single-logarithmic and 
therefore it is beyond control of the IREE with one mass scale. 
Finally, the third term, $\ln^2(M^2/\mu^2)$, does not depend on
$s$. It is usually dropped in the IREE with one mass scale. 

An expression 
for $\Psi_L$ similar to this one was obtained earlier in ref.\cite{flmm}
using a similar approach of writing an IREE with two infrared cut-offs. 
However, there are certain 
differences between our result and the one given in  
eq.~(28) of ref. \cite{flmm}. Besides the cut-offs $\mu$ and $M$,
Fadin {\em et al.} \cite{flmm} have another mass scale $m$ defined
such that $m<M$. Setting $m=M$ the result derived by Fadin {\em et
  al.} agrees with ours except for an over all factor $1/2$ which we
don't have. We believe that the origin of this disagreement could be
traced back to the use of Fadin {\em et  al.} of the axial
gauge. In fact, in this gauge, the DL contributions arise from fermion
self-energy diagrams. Then, it could be that the authors of ref.
\cite{flmm} give the one electron self-energy contribution rather
than the Sudakov form factor, which is the double of it.
There is another difference between our IREE for $F_L$ (see
eq.\ref{eqF}) and the corresponding equation in ref \cite{flmm}.
In the work of Fadin {\em et  al.} the rhs of the evolution equation
contains an extra logarithmic dependence on the fermion mass $m$.
The $m$ dependence comes from the fermion propagators. But, with DL
accuracy, one can write the propagators in terms of $\alpha$ and
$\beta$ as $(p_2 - k)^2 - m^2 = k^2 - 2p_2 k \approx -s\beta - k^2_{\perp} 
$. To obtain a log term from the $\beta$ integration one has to
require that $s\beta \gg k_\perp^2$. Then, this condition fixes the
lower limit of integration as $k_\perp^2$ and the remaining integral
is 
\begin{equation}
  \int_{\mu^2}^s\frac{dk_\perp^2}{k_\perp^2}\ln{(s/k_\perp^2)}
\end{equation}
 with no logarithmic contribution depending on $m$.
 
As a final remark, we would like to stress that the 
exponentiation of the DL contributions to  
the EW Sudakov form factor, $F_L$, takes place when both the initial
and the final state are not specified and summation over their weak
isospin is done. This means that, in contrast to QCD, such form
factors should be regarded as a theoretical object with rather limited
applications. In principle the off-shell version of $F_L$ could be
considered as an ingredient in the calculation of some more
complicated physical processes. 
   
\section{Acknowledgement}

The work is supported by grants CERN/2000/FIS/40131/   
and INTAS-97-30494.

\end{document}